\documentstyle[12pt,russian]{article}
\begin{document}
\setcounter{page}{28}
\begin{center}{\bf Geometrisation of electromagnetic field and topological
interpretation of quantum formalism}\end{center}
\begin{center}O.A.Olkhov\end{center}
\begin{center}{\it N.N.Semenov Institute of Chemical Physics, Russian 
Academy of Sciences}\end{center}
\begin{center}{\it Moscow, Russia, E-mail: olkhov@center.chph.ras.ru}
\end{center}

All numerous attempts of the electromagnetic field geometrisation were
based on concept "particle-field" (pointlike sources and extended field; see
for example [1-3]). But we had recently shown that the free Dirac field can be 
considered as a curved nonorientable closed connected 
time-space 4-manifold [4]. 
Its fundamental group consists of four global
gliding symmetries and the Minkowski space appears as the manifold 
universal covering space.
Taking this into account we now suggest a new approach to the electromagnetic
field geometrisation and this approach inevitably means a new
topological interpretation of the quantum mechanics mathematical formalism. 

Namely, we 
suppose that interacting Dirac and Maxwell fields can be 
considered as a single closed connected curved 4-manifold. 
Its fundamental group consists of 
four local gliding symmetries and its covering space has 
a more complex curved structure. Electric and magnetic fields appear within
such approach as components of the curvature tensor of the manifold 
covering space and
the Dirac spinors appear as basic functions of the manifold
fundamental group representation. 

Above concept differs from the one of
general theory of relativity by two main points: we geometrise not only
the field but we also geometrise the field sources and we
represent the field and the sources not as a curved riemannian 4-space but as a
4-manifold.

We start with the known equations for interecting clasical electromagnetic
and electron-positron fields [5]
$$i\gamma_1(\frac{\partial}{\partial x_1}+ieA_1)\psi-\sum_{\alpha =2}^4 
i\gamma_\alpha(\frac
{\partial}{\partial x_\alpha}+ieA_\alpha)\psi=m\psi,\eqno (1)$$
$$F_{kl}=\frac{\partial A_k}{\partial x_l}-\frac{\partial A_l}{\partial x_k},
\eqno (2)$$
$$\sum_{i=1}^4 \frac{\partial F_{ik}}{\partial x_i}=j_k,\quad
j_k=e\psi*\gamma_1\gamma_k\psi.\eqno (3)$$
Here $\hbar=c=1, x_1=t, x_2=x, x_3=y, x_4=z, F_{kl}$-tensor of electric and
magnetic fielfs, $A_k-4$-potential, $\gamma_k$-Dirac matrices, $\psi$-
Dirac spinor, $m$ and $e$-mass and charge of an electron.

It is shown that in (1) the expression
$\bigtriangledown_k=\partial/\partial x_k+ieA_k$
can be considered as the translation group generator into 
a conformal pseudoeuclidean 
4-space and $ieA_k$ appear within such approach as $\Gamma_k$
---the contraction of a riemannian connection  $\Gamma^q_{kp}$ of this space
$$\Gamma_k=\sum_p \Gamma^p_{kp}.$$

Multiplied by the reflection operators $\gamma_k$ [1] the $\bigtriangledown_k$
gives the representation of a local gliding symmetry group in this space
with the Dirac spinors as basic vecors of the represetation. All this gives the
opportunity to interpret (1) as the metric relation for a closed connected
nonorietable topological 4-manifold with the local 
gliding symmetry group as its
fundamental group and with a conformal pseudoeuclidean space as its
universal covering space (two-dimentional analog of such manifold is the
combination of a torus and the Klein bottle [6]).

Now we use a relation between a riemannian connection  
and a riemannian curvature tenor $R^q_{lk,i}$ [7]
$$R^q_{lk,i}=\frac{\partial \Gamma^q_{li}}{\partial x_k}-
\frac{\partial \Gamma^q_{ki}}{\partial x_l}+\Gamma^q_{kp}\Gamma^p_{li}-
\Gamma^q_{lp}\Gamma^p_{ki}.$$
After contraction over $q$ and $l$ we obtain
$$R^o_{ik}=\sum_q R^q_{ik,q}=\frac{\partial \Gamma_i}{\partial x_k}-
\frac{\partial \Gamma_k}{\partial x_i}.\eqno (4)$$

By comparison (2) and (4) we can write down the equations (1-3)
using only geometrical notations
$$i\gamma_1\bigtriangledown_1\psi-\sum^4_{\alpha=2} i\gamma_\alpha 
\bigtriangledown_\alpha \psi=
m\psi,\eqno (5)$$
$$\sum^4_{i=1} \frac{\partial R^o_{ik}}{\partial x_i}=j_k,\eqno (6)$$
where $R^o_{ik}$ has the form (4).

Finally, we have the following geometrical 
interpretation of electromagnetic field.

\noindent 1. Electromagnetic field and its sources (electron-positron field) 
can be considered as a single 4-manifold
which fundamental group is generated by four local gliding symmetries.\\
2. Universal covering space of this manifold is the conformal
pseudoeuclidean space.\\
3. The 4-potentials $A_k$ is defined by the connections 
of this space $\Gamma_k$ ($ieA_k=\Gamma_k$).\\
4. The electric and magnetic field components are defined by the
components  of the covering space curvature tensor $R_{ik}$
($ieF_{ik}=R_{ik}$).\\
5. The Dirac spinors appear as basic functions for  the manifold fundamental 
group representation.

One comment in conclusion. We see that replacing a "wave-particle" by a
space-time manifold does not mean "more determinism" for the quantum
object description. And we also see that the above topological approach
does not introduce any hidden variables and does not therefore contradict
the Bell and von Neumann theorems [8,9].\\

\noindent 1. Weyl H. Gravitation und Electrizitat. 
Berlin, Sitzber.Preuss.Akad.Wiss, 1918\\
2. Konopleva N.P., Sokolic H.A., Nucl.Phys. 1965. V.72. P.667\\
3. Daniel M., Viallet C.M., Rev.Mod,Phys. 1980. V.52. P.175\\
4. Olkhov O.A., e-print quant-ph/0101137, in 
Proceedings of the 7th International Symposium on Particles, Strings and
Cosmology, Lake Tahoe, California, 10-16 December 1999. Singapure-New Jersey-
Hong Kong.: World Scientific, 2000, P.160.\\
5. Schweber S., An introduction to relativistic quantum field theory, Row,
Peterson and Co., Evanstone, Ill., Elmford, N.Y. 1961\\
6. Coxeter H.S.M., Introduction to geometry, John Willey and Sons, Inc.,
N.Y.-London, 1961.\\
7. Dirac P.A.M., General theory of relativity, John Wiley and Sons, N.Y.-
Sydney-Toronto, 1975\\
8. Bell I.S., Rev.Mod.Phys. 1966. V.38. P.447\\
9. von J.V.Neumann., Mathematische grundlagen der quantenmechanik,
Verlag von Julius Springer, Berlin, 1932.

\end{document}